\documentclass[a4paper,11pt]{article}
\pdfoutput=1 

\usepackage{jinstpub} 

\title{First results with the ANET Compact Thermal Neutron Collimator}

\author[a,b,1]{O. Sans-Planell,\note{Corresponding author.}}
\author[a,b]{M. Costa,}
\author[a,b]{E. Durisi,}
\author[a,b]{E. Mafucci,}
\author[a,b]{L. Menzio,}
\author[a,b]{V. Monti,}
\author[a,b]{L. Visca,} 

\author[c,d]{F. Grazzi,}

\author[e]{R. Bedogni,}

\author[f,g]{S. Altieri}

\affiliation[a]{Universita degli Studi di Torino,\\Via Pietro Giuria 1, 10125, Torino, Italy}
\affiliation[b]{INFN, Sezione di Torino,\\Via Pietro Giuria 1, 10125, Torino, Italy}
\affiliation[c]{Consiglio Nazionale delle Ricerche, Istituto dei Sistemi Complessi,\\ Via Madonna del Piano 10, I-50019 Sesto Fiorentino, Italy}
\affiliation[d]{INFN, Sezione di Firenze,\\Via Sansone 1, 50019, Sesto Fiorentino, Italy}
\affiliation[e]{INFN, Laboratori Nazionali di Frascati,\\Via Enrico Fermi 40, 00044, Frascati, Italy}
\affiliation[f]{Università degli Studi di Pavia,\\Via Agostino Bassi, 6, 27100 Pavia, Italy}
\affiliation[g]{INFN, Sezione di Pavia,\\Via Agostino Bassi, 6, 27100 Pavia, Italy}
\emailAdd{oriol.sansplanell@to.infn.it}

\abstract{This paper presents the first determination of the spatial resolution of the ANET Compact Neutron Collimator, obtained with a measuring campaign at the LENA Mark-II TRIGA reactor in Pavia. This novel collimator consists of a sequence of collimating and absorbing channels organised in a chessboard-like geometry. It has a scalable structure both in length and in the field of view. It is characterized by an elevated collimation power within a limited length. Its scalability and compactness are added values with respect to traditional collimating system. The prototype tested in this article is composed of 4 concatenated stages, each 100$mm$ long, with a channel width of 2.5$mm$, delivering a nominal L/D factor of 160. This measuring campaign illustrates the use of the ANET collimator and its potential application in neutron imaging for facilities with small or medium size neutron sources.}


\arxivnumber{1234.56789} 

\begin{document}
\maketitle
\flushbottom

\section{Introduction}

This paper is dedicated to the study and characterisation of a Compact Neutron Collimator (CNC) developed within the framework of the ANET project \cite{a,c}. The measurement campaign has been performed at the LENA 250$kW$ Mark-II TRIGA reactor in Pavia (Italy). The thermal neutron beam provided by the LENA facility is poorly collimated and thus it can be used as a test-bench to verify the ANET CNC collimation performances and its possible application for neutron imaging. To measure the image resolution, two reference test objects have been used, i.e. a gadolinium Siemens star and a gadolinium bar pattern  developed at the Paul Scherrer Institute \cite{b}. These two devices allow to determine the image resolution in a range between 25$\mu m$ and 1000$\mu m$. In the following sections the measurement set-up and methodology are described and a demonstration of the impact of the ANET CNC in terms of image resolution is illustrated.

\section{Experimental set-up and methodology}

\begin{figure}
    \centering
    \includegraphics[width=1\textwidth]{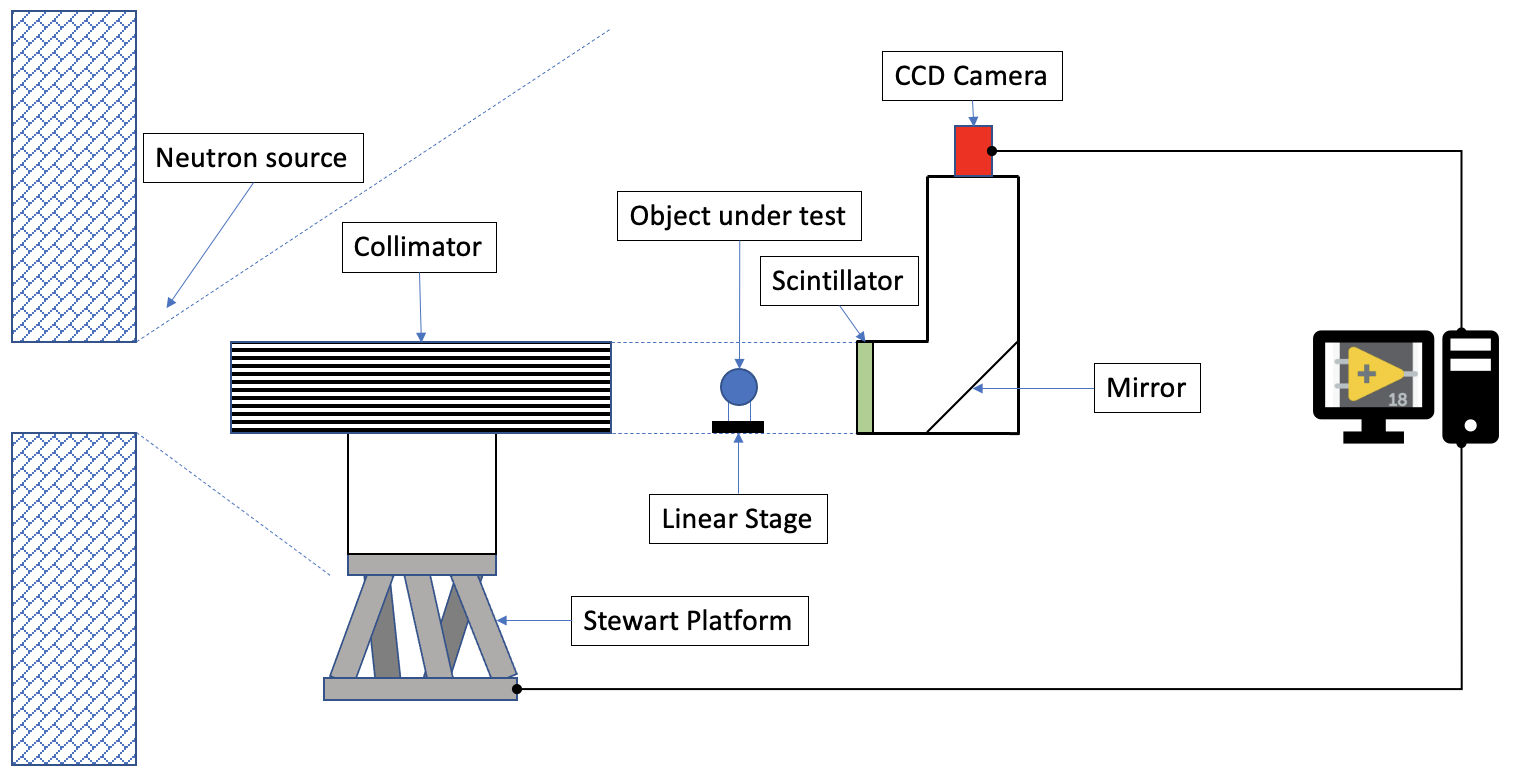}
    \caption{Schematics of the neutron imaging setup with the ANET collimator.}
    \label{fig:setup}
\end{figure}

The experimental set-up used is composed of 4 main stages (figure \ref{fig:setup}): the LENA thermal neutron source, the ANET CNC coupled with a Physik Instrumente Stewart platform, the test object and a commercial neutron camera ensemble  \cite{c,e,b,d}. \newline

\begin{figure}
    \centering
    \includegraphics[width=1\textwidth]{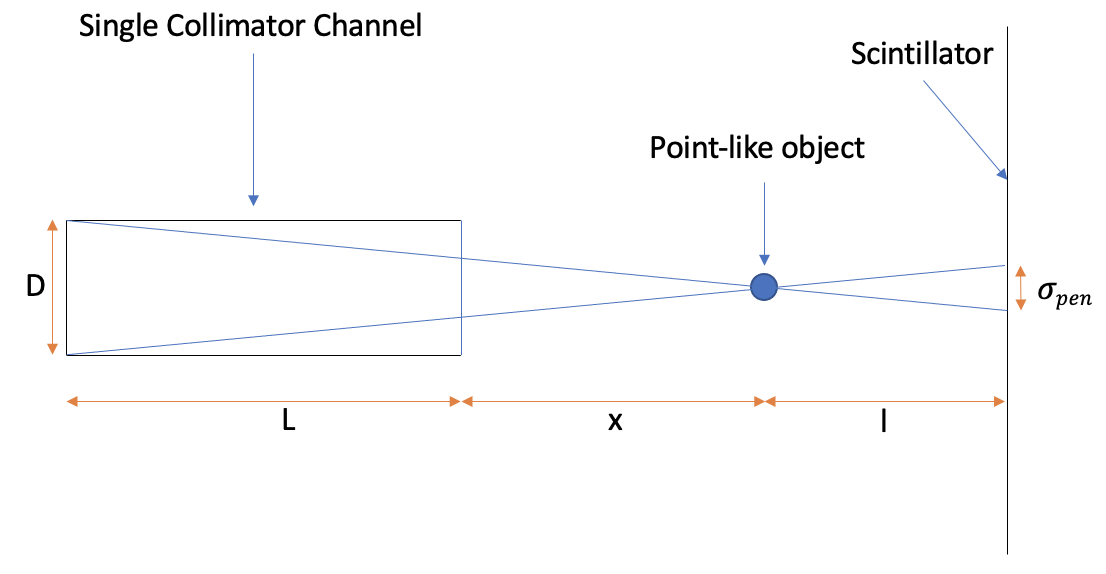}
    \caption{Geometrical representation of a single collimation channel. For the definition of the single quantities refer to the text.}
    \label{fig:schematics-ld}
\end{figure}

In figure \ref{fig:schematics-ld} a single collimation channel is shown and the relevant variables are defined:
\begin{itemize}
    \item D: Diameter of the collimator channel.
    \item L: Length of the collimator channel.
    \item x: Distance of the object from the collimator.
    \item l: Distance of the object from the scintillator.
    \item $\sigma_{pen}$: geometrical resolution due to the penumbra blurring.
\end{itemize}

The ANET CNC, as explained in ref. \cite{c}, is a multi-channel device with air collimation channels alternated to highly absorbing channels. It is operated following a dynamic acquisition protocol to deliver a clean neutron radiography, whithout the possible artefacts due to the discrete structure of the collimator. To this end, the CNC is mounted on a Stewart platform, capable of moving it through 6 independent degrees of freedom with a translational precision of $10\mu m$  and a rotational precision of $12.5\mu rad$. These levels of accuracy are required during the preliminary alignment procedure and the main image acquisition, as explained in the following section. \newline
The sample object, in front of the collimator, is moved in and out of the field of view by mean of a small linear stage. The image obtained when the sample object is out of the beam defines the so-called "Open Beam Image".\newline
The neutron camera is a commercial device \cite{d}, composed by a 14-bit Sony CCD coupled to a 400$\mu m$ scintillator screen made of ZnS(Cu) covered with ${}^{6}$LiF at a mass rate 2:1. The camera is cooled at 20 degrees below the room temperature to minimise the electronic noise.

Any neutron radiography ($I$) should be properly normalised  using an open beam image ($I_{OB}$) and a dark field image ($I_{DF}$) under the same exposure conditions. The quantities $I$, $I_{OB}$ and $I_{DF}$ are expressed in pixel counts.
The normalized neutron radiography $F_{rad}$ is defined as follows:
\begin{equation}
    F_{rad} = \frac{I - I_{DF}}{I_{OB} - I_{DF}}
    \label{eq:frad}
\end{equation}

\subsection{The CNC alignment procedure}
\begin{figure}[htb]
    \centering
    \includegraphics[width=0.8\textwidth]{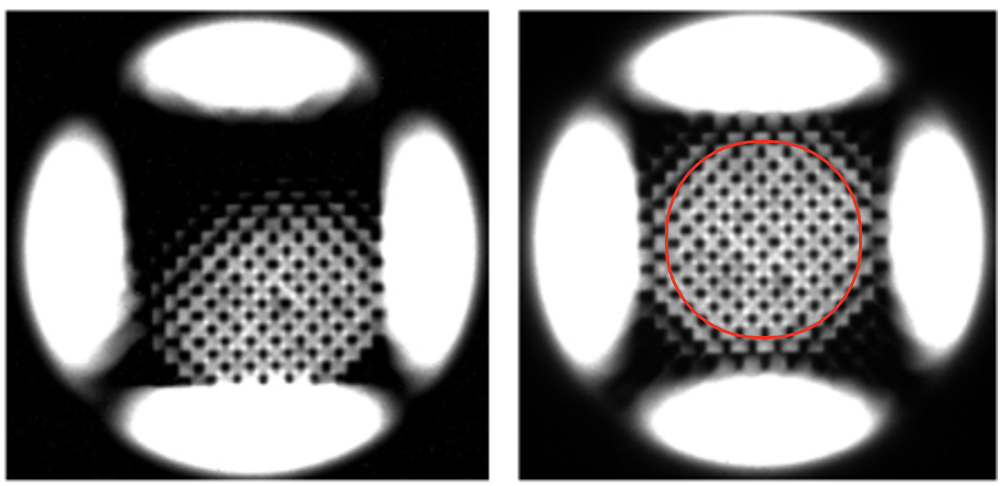}
    \caption{Neutron image of the collimator before the alignment (left) and after the alignment (right).The red circle indicates the effective field of view.}
    \label{fig:align}
\end{figure}

The ANET CNC prototype has a field of view of $50mm x 50mm$ and a nominal L/D factor of 160. It is 400 mm long, and each collimation channel is $2.5mm$ wide, corresponding to a maximum neutron angular divergence of 0.36 degree (or $6\cdot 10^{-3} rad$). Given these constraints, the alignment of the collimator w.r.t. the beam axis is crucial.\newline
The PI Stewart Platform specifications allow for the angular fine tuning of the collimator alignment with the required precision. The static image of the collimator before and after the alignment is shown in figure \ref{fig:align}.
The image is taken with the neutron camera described in the previous section. The chess-board image reflects the collimator structure. The image after the alignment shows a good level of uniformity in the inner part while the effects of the conical shape of the Pavia LENA beam are more visible at the edges and the corners. This results in an effective field of view with 40 mm diameter that is shown as a circle in figure \ref{fig:align} (right).


\subsection{The CNC method of operation}
\begin{figure}[htb]
    \centering
    \includegraphics[width=1\textwidth]{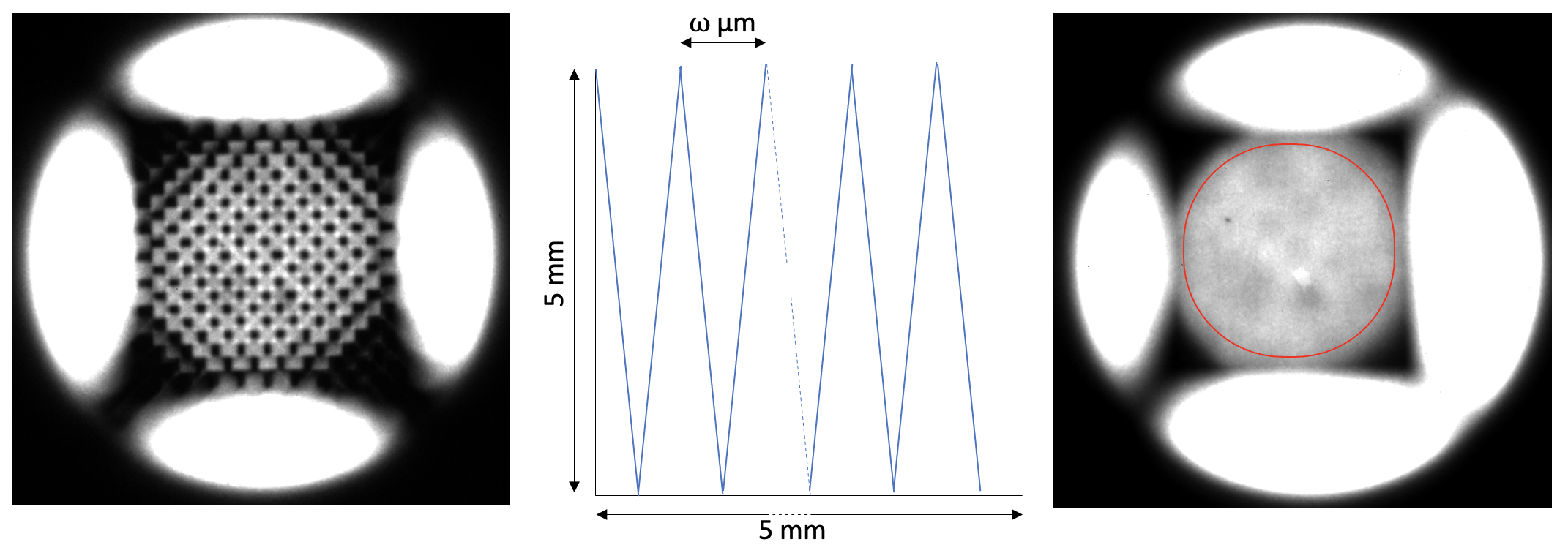}
    \caption{Neutron image of the collimator (Left). Example of dynamic pattern (centre). Image of the open beam with the dynamic pattern (right). The red circle indicates the effective field of view.}
    \label{fig:pattern}
\end{figure}

In order to properly use the ANET CNC avoiding the artefacts induced by the chessboard structure in the final image a dynamic image acquisition approach has been developed. 
In order to optimise the image quality, whilst minimising the data processing, a long exposure radiography with the collimator moving continuously is taken.
The time of exposure is chosen to maximize, in the region of interest, the CCD pixel counts without reaching pixel saturation. This, of course, depends on the neutron source intensity.
In the Pavia conditions, the exposure time has been set to 900$s$ along which the Stewart Platform moved the CNC following a specific pattern confined within 5 x 5 $mm^2$ area, equivalent to two air channels and two absorbing rods. Extensive tests using different pattern geometries have been done. The "saw" profile shown in figure \ref{fig:pattern} turned out to be the optimal one. In figure \ref{fig:pattern}, $\omega$ is equivalent to the scan range (5$mm$) divided by half of the number of divisions. This results in a value of 47.6$\mu m$, smaller than the size of the projected pixel on the scintillator, which is 47.9$\mu m$. The overall result (figure \ref{fig:pattern} right) is a smooth, pattern-less image. \newline

\begin{figure}[htb]
    \centering
    \includegraphics[width=.6\textwidth]{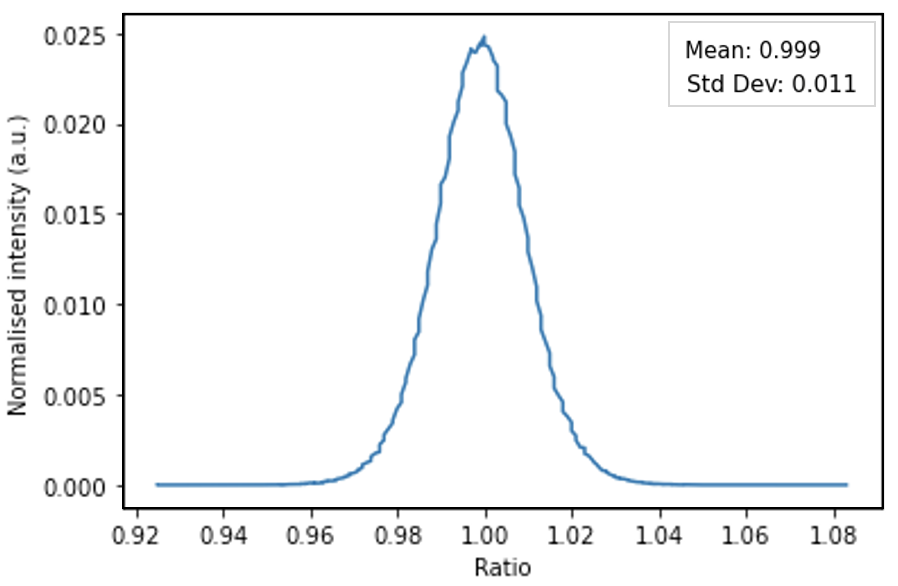}
    \caption{Histogram of the ratio of two open beam images.}
    \label{fig:histogram}
\end{figure}

The uniformity of the OB image is an important factor as it determines the reproducibility and quality of the normalisation procedure described in equation \ref{eq:frad}. This can be evaluated by calculating the ratio of different OB images and extracting the intensity distribution of the pixels within the field of view. The resulting histogram in figure \ref{fig:histogram} shows a mean of 0.999 with a standard deviation of 0.011, implying a uniformity and repeatability within $\pm1\%$.

\section{Determination of the ANET CNC resolution}

\begin{figure}[htb]
    \centering
    \includegraphics[width=.9\textwidth]{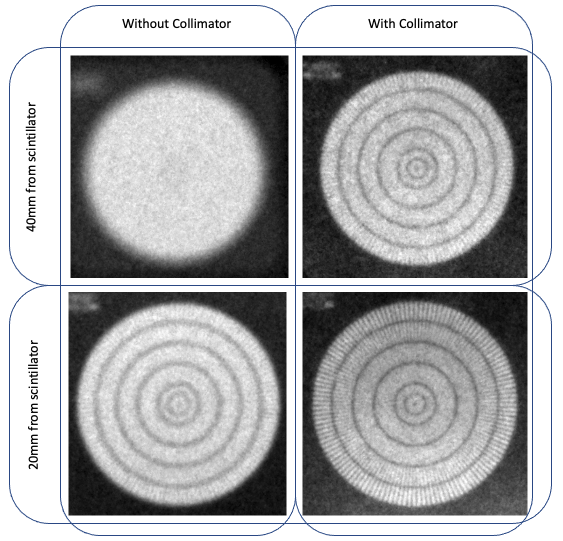}
    \caption{Table showing the normalised measurements of the Siemens star with and without the collimator at 40$mm$ and 20$mm$ from the scintillator respectively.}
    \label{fig:siemens-with-and-out}
\end{figure}

\begin{figure}[htb]
    \centering
    \includegraphics[width=0.8\textwidth]{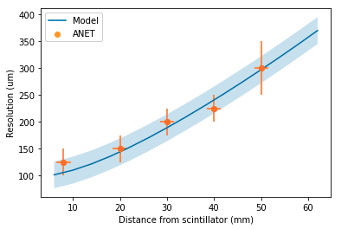}
    \caption{Comparison between the resolution obtained with the ANET collimator and that expected from the theoretical model.}
    \label{fig:ANET-theory}
\end{figure}

To evaluate the system spatial resolution, a set of Siemens star and bar pattern has been used \cite{b}. These instruments cover a resolution ranges of 25-250$\mu m$ and 50-1000$\mu m$ respectively.
Keeping the collimator and the scintillator at a distance of 100$mm$ from each other, different measurements were taken with the Siemens star and the bar pattern at variable distances from the scintillator. The measurements were repeated with and without the ANET collimator. The impact of the ANET collimator can be appreciated in figure \ref{fig:siemens-with-and-out}. 
The radiography images taken with the collimator shows an evident improvement in the spatial resolution with respect to the same image taken without the ANET CNC. \newline
To be more quantitative and to describe the behaviour of the spatial resolution as a function of the distance $l$ from the scintillator, it is worth to extract the relation between the geometrical resolution $\sigma_{pen}$ and the quantities $x$ and $l$ defined in figure \ref{fig:schematics-ld}. From geometrical considerations (see figure \ref{fig:schematics-ld}) it can be easily derived:
\begin{equation}
    \sigma_{pen} = \frac{l D}{L + x}
    \label{eq:resolution}
\end{equation}
The total resolution $\sigma_{total}$ can be calculated through \cite{b}: 
\begin{equation}
\sigma_{total}=\sqrt{\sigma^{2}_{pen}+\sigma_{det}^2}
\label{eq:sigmatot}
\end{equation}
Where $\sigma_{det}$ is the contribution due to the neutron camera detector ensemble, which is related to the inverse of the Nyquist frequency, defined as the minimum measurable separation between two lines and numerically equivalent to the double of the pixel size projected at the scintillator plane. In the case of the neutron camera used for this experiment $\sigma_{det} = 96 \mu m$\newline
The total resolution appreciable in a measurement is then a combination of the beam collimation quality and the contribution of the detecting system. 
\begin{figure}[]
    \centering
    \includegraphics[width=.8\textwidth]{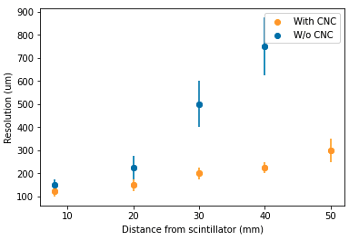}
    \caption{Comparison between the resolution obtained with and without the ANET collimator as a function of the distance of the Siemens star from the scintillator.}
    \label{fig:ANET-LENA}
\end{figure}
The comparison between the total resolution measured at LENA in Pavia with and without the ANET collimator at various distances between the test object and the scintillator is shown in figure \ref{fig:ANET-LENA}.\newline
The image resolution improvement due to the ANET collimator is mighty and it becomes more relevant with increasing distance.
The total resolution values with ANET CNC range between 100 and 300 microns.\newline
The measured points behaviour can be described by the simple geometrical model derived from equation \ref{eq:sigmatot} The error associated with the model is half a pixel, as the resolution true value may vary within this range. The comparison is shown in Figure \ref{fig:ANET-theory}: the theoretical curve and the data points are in good agreement.

\section{Conclusions}
This paper shows the first results obtained with the ANET CNC compact collimator prototype during the measuring campaign at the LENA Mark-II TRIGA reactor, during summer 2021. \newline
The measurement serves as a proof of concept that the ANET multi-channel 2D collimator employing a dynamic pattern acquisition technique can serve as a core part of an imaging setup\newline
The resolution measurements follow the theoretical expectations and  demonstrates the ANET CNC capability to improve beam lines with limited collimation power. \newline
Future campaigns with higher intensities and different neutron beam facilities will be useful to demonstrate the ANET collimator portability and performances.

\acknowledgments

We acknowledge INFN CSN5 for funding the project and the INFN Frascati Mechanical Laboratory for their support. \newline
We acknowledge as well the LENA personnel for their support and collaboration.



\begin{thebibliography}{99}

\bibitem{a}
PCT Patent n. PCT/IB2021/053856  Collimatore

\bibitem{c}
Bedogni, R., et al. (2021). Design of a novel compact neutron collimator. Journal of Instrumentation, 16(08), P08055.

\bibitem{b}
Kaestner, A. P., et al. (2017). Samples to determine the resolution of neutron radiography and tomography. Physics Procedia, 88, 258-265.

\bibitem{d}
Hewat, A. W. Inexpensive Neutron Imaging Cameras using CCDs for Astronomy. Physics Procedia, 2015, 69: 185-188.

\bibitem{e}
https://www.physikinstrumente.com/en/technology/parallel-kinematics/multi-axis-positioners/ visited on 10-August-2021.





\end{thebibliography}
\end{document}